\documentclass[twocolumn,showpacs,aps,prd,superscriptaddress,amsmath,amssymb,floatfix]{revtex4}

\usepackage{graphicx}
\usepackage{dcolumn}
\usepackage{bm}
\usepackage{epsfig}
\usepackage{afterpage}
\usepackage{psfrag}
\usepackage{url}
\usepackage{multirow}

\input{pubboard/babarsym.tex}

\newcommand{\KstarzbEightNineTwo}{\ensuremath{\Kbar^{*0}(892)}\xspace}
\newcommand{\figureref}[1]       {Fig.~\ref{fig:#1}}

\newcommand{\KstarII}            {\ensuremath{\Kbar^{*0}_{0}(1430)}\xspace}

\newcommand{\Btokstark}          {\ensuremath{\Bp \to \KstarzbEightNineTwo \Kp}}
\newcommand{\BBpairs}            {\ensuremath{232}}
\newcommand{\sigYield}           {\ensuremath{25\pm13\,[stat]\,\pm7\,[syst]}}
\newcommand{\Btokstarktokpi}     {\ensuremath{\Bp \to \KstarzbEightNineTwo (\to \Km \pip) \Kp}}
\newcommand{\BRmeas}             {\ensuremath{(0.6\pm0.3\,[stat]\,\pm0.2\,[syst])\times10^{-6}}}

\newcommand{\BRul}               {\ensuremath{1.1\times10^{-6}}}
\newcommand{\SM}                 {Standard Model}
\newcommand{\BtoPhiKsZero}       {\ensuremath{\Bz \to \phi \KS}}
\newcommand{\BtokstarIIk}        {\ensuremath{\Bp \to \KstarII \Kp}}
\newcommand{\BRulII}             {\ensuremath{2.2\times10^{-6}}}

\newcommand{\GuoPred}            {\ensuremath{0.31\times10^{-6}}}
\newcommand{\thryLL}             {\ensuremath{(0.46^{+0.06}_{-0.07})\times10^{-6}}}

\newcommand{\cleoRes}            {\ensuremath{5.3\times10^{-6}}}
\newcommand{\jBtoKKpi}           {\ensuremath{\Bp \to \Kp \Km \pip}}
\newcommand{\BtoKstKGeneric}     {\ensuremath{\B \to \Kstar \kaon}}
\newcommand{\btodssbar}          {\ensuremath{\b \to \d \s \sbar}}
\newcommand{\btod}               {\ensuremath{\b \to \d}}
\newcommand{\deltaS}             {\ensuremath{\Delta S_{\phi \KS}}}
\newcommand{\btos}               {\ensuremath{\b \to \s}}
\newcommand{\tesla}              {\ensuremath{\rm \,T}\xspace}
\newcommand{\BBpairsErr}         {\ensuremath{3}}
\newcommand{\LUMI}               {\ensuremath{211}}
\newcommand{\OffResLumi}         {\ensuremath{22}}
\newcommand{\jTa}                {\ensuremath{\theta_{T}}\xspace}
\newcommand{\jT}                 {\ensuremath{\cos{\jTa}}\xspace}
\newcommand{\jF}                 {\ensuremath{{\cal F}}\xspace}
\newcommand{\jFb}                {{\boldmath \ensuremath{{\cal F}}}\xspace}
\newcommand{\jI}                 {\ensuremath{m_{K \pi}}\xspace}
\newcommand{\jIb}                {{\boldmath \ensuremath{m_{K \pi}}}\xspace}
\newcommand{\jHa}                {\ensuremath{\theta_{H}}\xspace}
\newcommand{\sigEff}             {\ensuremath{26\%}}
\newcommand{\jH}                 {\ensuremath{\cos{\jHa}}\xspace}
\newcommand{\jHb}                {{\boldmath \ensuremath{\cos{\jHa}}}\xspace}
\newcommand{\btoc}               {\ensuremath{\b \to \c}}
\newcommand{\BtoPhiK}            {\ensuremath{\Bp \to \phi \Kp}}
\newcommand{\BtoKKK}             {\ensuremath{\Bp \to \Kp \Km \Kp}}
\newcommand{\BtoRhoK}            {\ensuremath{\Bp \to \rho \Kp}}
\newcommand{\BtoPhiKYld}         {\ensuremath{21\pm2}}
\newcommand{\BtoKKKYld}          {\ensuremath{23\pm2}}
\newcommand{\BtoRhoYld}          {\ensuremath{4\pm1}}
\newcommand{\jKKpi}              {\ensuremath{\Kp \Km \pip}}
\newcommand{\nonKKpBBsub}        {\ensuremath{3}}
\newcommand{\BtokstarIIktokpi}   {\ensuremath{\Bp \to \KstarII (\to \Km \pip) \Kp}}
\newcommand{\jKpi}               {\ensuremath{\Km \pip}}
\newcommand{\jM}                 {\ensuremath{m_{\rm ES}}\xspace}
\newcommand{\jMb}                {{\boldmath \ensuremath{m_{\rm ES}}}\xspace}
\newcommand{\jD}                 {\ensuremath{\Delta E}\xspace}
\newcommand{\jDb}                {{\boldmath \ensuremath{\Delta E}}\xspace}

\newcommand{\KstarIItokpi}       {\ensuremath{\KstarII \to \Km \pip}}
\newcommand{\BtokstarIIkPresent} {\ensuremath{5^{+14}_{-5}}}
\newcommand{\BtokstarIIkBias}    {\ensuremath{2^{+6}_{-2}}}
\newcommand{\Like}               {\ensuremath{{\cal L}}\xspace}
\newcommand{\efc}                {\hspace{1ex}}
\newcommand{\ARGUS}              {\mbox{\scshape ARGUS}}
\newcommand{\expFunc}            {\ensuremath{\exp(P) + C}}

\newcommand{\jimBtoDpiCalib}     {\ensuremath{\Bp \to \Dzb \pip}}

\newcommand{\effSystErr}         {\ensuremath{\pm13 \%}}

\newcommand{\jmrTwo}[1]          {\multirow{2}{*}{#1}}
\newcommand{\numEvts}            {\ensuremath{38,690}}
\newcommand{\rawYield}           {\ensuremath{30\pm13\,[stat]\,\pm3\,[syst]}}
\newcommand{\BBbkgdSub}          {\ensuremath{5^{+7}_{-4}\,[syst]}}

\newcommand{\BRmeasMoreDet}      {\ensuremath{(0.6\pm0.3\,[stat]\,\pm0.2\,[syst])\times10^{-6}}}
\newcommand{\signif}             {\ensuremath{3.1 \sigma}}
\newcommand{\signifWithSyst}     {\ensuremath{1.6 \sigma}}

\newcommand{\BtoPhipi}           {\ensuremath{\Bp \to \phi \pip}}
\newcommand{\deltaSBound}        {\ensuremath{0.11}}

\newcommand{\ChiangPred}         {\ensuremath{0.5\times10^{-6}}}

\begin{document}

\preprint{\babar-PUB-07/004}
\preprint{SLAC-PUB-12546}
\preprint{arXiv:0706.1059 [hep-ex]}
\preprint{BAD \#1660 Version 14}

\begin{flushleft}
\babar-PUB-07/004\\
SLAC-PUB-12546\\
\end{flushleft}

\title{Search for the decay {\boldmath $B^{+} \rightarrow \kern 0.20em\overline{\kern -0.20em K}{}^{*0}(892) K^{+}$}}

%
\author{B.~Aubert}
\author{M.~Bona}
\author{D.~Boutigny}
\author{Y.~Karyotakis}
\author{J.~P.~Lees}
\author{V.~Poireau}
\author{X.~Prudent}
\author{V.~Tisserand}
\author{A.~Zghiche}
\affiliation{Laboratoire de Physique des Particules, IN2P3/CNRS et Universit\'e de Savoie, F-74941 Annecy-Le-Vieux, France }
\author{J.~Garra~Tico}
\author{E.~Grauges}
\affiliation{Universitat de Barcelona, Facultat de Fisica, Departament ECM, E-08028 Barcelona, Spain }
% ...
\collaboration{The \babar\ Collaboration}
\noaffiliation

\date{\today}

\begin{abstract}

We report on a search for the process $B^{+} \rightarrow
\kern 0.20em\overline{\kern -0.20em K}{}^{*0}(892) K^{+}$
using $232 \times 10^{6}$ $\Upsilon(4S) \rightarrow B
\kern 0.18em\overline{\kern -0.18em B}{}$ decays collected
with the \mbox{\slshape B\kern-0.1em{\small A}\kern-0.1em
B\kern-0.1em{\small A\kern-0.20em R}} detector at the
PEP-II asymmetric-energy $B$ Factory at SLAC.
From a signal yield of $25 \pm 13\,[stat]\,\pm 7\,[syst]$
$B^{+} \rightarrow \kern 0.20em\overline{\kern -0.20em K}{}^{*0}(892)
(\rightarrow K^{-} \pi^{+}) K^{+}$ events,
we place an upper limit on the branching fraction
${\cal B}(B^{+} \rightarrow
\kern 0.20em\overline{\kern -0.20em K}{}^{*0}(892) K^{+})$
of $1.1 \times 10^{-6}$, at the $90 \%$ confidence level.

\end{abstract}


\maketitle

We present a measurement of the branching fraction
${\cal B}(\Btokstark)$ based exclusively on \Bp\ decays
to the final state \jKKpi.
Charge conjugate states are assumed throughout.
In the \SM\ (SM), \BtoKstKGeneric\ decays are dominated by
\btodssbar\ gluonic penguin diagrams (see Figure 1(a) in~\cite{Aubert:2006wu};
for the charged decay the spectator \dbar\ is replaced
with \ubar).
Such transitions provide a valuable tool with which to test
the quark-flavor sector of the SM
(see, for example,~\cite{Fleischer:2004rnAndFleischer:2005gj,Chiang:2003pm,Guo:2006uq}).
The mode \Btokstark\ is also relevant for the interpretation
of the time dependent \CP\ asymmetry obtained with the
\BtoPhiKsZero\ mode.
To leading order the \CP\ asymmetry equals \stwob\ for this mode,
where $\beta$ is the Unitarity Triangle (UT) angle.
However, sub-dominant amplitudes proportional to $V_{ub}^{*} V_{us}$
could produce a deviation \deltaS\ from \stwob.
Exploiting SU(3) flavor symmetry and combining measured rates
for relevant \btos\ and \btod\ processes (including \Btokstark),
a method is introduced in~\cite{Grossman:2003qp} to place a
bound on \deltaS.
Measurements yielding a significant deviation in excess of such a
bound would be a strong indication of physics beyond the SM.
Furthermore, \Btokstark\ is one of several charmless
decays
that can be used, together with U-spin symmetry, to extract
the UT angle $\gamma$~\cite{Soni:2006vi}.

Theoretical predictions for $\BR(\Btokstark)$ include
$\BR(\Btokstark) > \thryLL$~\cite{Fleischer:2004rnAndFleischer:2005gj}
and $\BR(\Btokstark) \approx \ChiangPred$~\cite{Chiang:2003pm}---both
using SU(3) flavor symmetry and experimental information
for charmless \B\ decays, and $\BR(\Btokstark) \approx
\GuoPred$---using perturbative QCD factorization~\cite{Guo:2006uq}.
Prior to the analysis presented here, the only experimental
limit placed on $\BR(\Btokstark)$ was that presented by the
CLEO collaboration at the $90 \%$ confidence
level (CL)~\cite{Jessop:2000bv}: $\BR(\Btokstark) < \cleoRes$.

The data used in this analysis were collected with the
\babar\ detector~\cite{Aubert:2001tu} at the \pep2\ asymmetric-energy
\epem\ storage ring at the Stanford Linear Accelerator Center.
Charged particle trajectories are measured by a five-layer double-sided
silicon vertex tracker and a 40-layer drift chamber located within a
$1.5 \tesla$ axial magnetic field.
Charged hadrons are identified by combining energy loss information from
tracking (\dedx) with the measurements from a ring-imaging Cherenkov detector.
Photons and electrons are detected by a CsI(Tl) crystal electromagnetic
calorimeter.
The magnet's flux return is instrumented for muon and
neutral hadron identification.

The data sample consists of $(\BBpairs \pm \BBpairsErr) \times 10^{6}$ \BB\ pairs
collected at the \FourS resonance (on-resonance data), corresponding to an integrated
luminosity of $\LUMI \invfb$.
It is assumed that neutral and charged \B\ meson pairs are produced
in equal numbers~\cite{Aubert:2004ur}.
In addition, $\OffResLumi \invfb$ of data collected $40 \mev$ below the
\FourS\ resonance (off-resonance data) are used for background studies.

\B\ meson candidates are reconstructed from three charged tracks.
The charged tracks are required to have at least $12$ hits in the drift
chamber and a transverse momentum greater than $0.1 \gevc$.
They are fitted to a common vertex; momentum must be conserved at this vertex.
Two of the tracks must have opposite charge and a signal in the tracking and
Cherenkov detectors that is consistent with that of a kaon.
We remove tracks that pass electron selection criteria based on
\dedx\ and calorimeter information.

We perform full detector Monte Carlo (MC) simulations equivalent to
$2.4 \times 10^{5}$ signal \Btokstarktokpi\ decays.
For background studies $1.0 \invab$ of generic \BB decays
are simulated, as are over $100$ exclusive \B\ meson decay modes
($\sim 10^{4}-10^{6}$ events/mode),
approximately half of which are charmless.
MC samples are generated with EvtGen~\cite{Lange:2001uf}, while
the detector response is simulated with GEANT4~\cite{Agostinelli:2002hh}.
All simulated events are reconstructed in the same manner as data.
Off-resonance data are used to measure the properties of the light
quark continuum decays, $\epem \to \qqbar$ ($q = u$, $d$, $s$, $c$).

For correctly reconstructed signal events,
$\DeltaE=E^{*}_{B}-\sqrt{s}/2$ peaks at zero, while
$\jM=\sqrt{(s/2+{\bf p}_{0}\cdot
{\bf p}_{B})^{2}/E^{2}_{0}-{\bf p}^{2}_{B}}$ peaks at
the \B\ mass.
The resolutions of these largely uncorrelated kinematic variables, for
signal events, are $\approx 20 \mev$ and $\approx 2.5 \mevcc$,
respectively.
$E^{*}_{B}$ is the \B\ meson candidate energy in the center-of-mass (CM)
frame, $E_{0}$ and $\sqrt{s}$ are the total energies of the $\epem$
system in the laboratory and CM frames, respectively, and
${\bf p}_{0}$ and ${\bf p}_{B}$ are the three-momenta of the $\epem$
system and the \B\ meson candidate in the laboratory frame.
The distributions for continuum events are slowly varying.
For \B\ meson decays in which particle misidentification occurs,
the $\DeltaE$ peak is shifted by $500$ or more $\mev$.
Events are selected with $5.22<\jM<5.29 \gevcc$ and
$|\DeltaE|<0.1 \gev$.
The \jD\ restriction
helps to remove background
from two- and four-body \B\ meson decays at a small cost to signal
efficiency.

Continuum quark-antiquark production is the dominant background.
To suppress it, we select only those events where the angle
\jTa\ in the CM frame between the thrust axis of the
\B\ meson candidate and the thrust axis of the rest of the event
satisfies $|\jT| < 0.9$.
For continuum events, which tend to be jet-like in the CM frame,
the distribution of $|\jT|$ is strongly peaked toward unity
whereas the distribution is uniform for signal events in which
little kinetic energy is available in the CM frame.
The number of continuum background events present per signal
event is reduced by a factor of approximately two with the
application of the $|\jT|$ cut.
We also construct a Fisher discriminant~\cite{Aubert:2005ce} \jF,
a linear combination of five variables: the zeroth and second angular
moments of the energy flow---excluding the \B\ candidate---about the
\B\ thrust axis; the absolute value of the cosine of the angle between the
momentum vector of the reconstructed \B\ candidate and the beam direction;
the absolute value of the cosine of the angle between the thrust
axis of the reconstructed \B\ candidate and the beam direction; and
the output of a multivariate, non-linear \B\ meson candidate flavor
tagging algorithm~\cite{Aubert:2002ic}.
The Fisher coefficients are obtained from samples of off-resonance
data and \jBtoKKpi\ MC.
A loose cut of $|\jF| < 3.0$ is applied.
This cut eliminates a negligible fraction of
signal and background events and is applied only
to define a range for the fit (see below).

Further discrimination between signal and continuum
background is achieved by utilizing the variables
\jI\ and \jH.
The invariant mass of the \KstarzbEightNineTwo\ candidate, \jI, is
restricted to $0.744<\jI<1.048 \gevcc$.
We also require that the cosine of the helicity angle,
\jHa, is less than $0.9$,
where \jHa\ is defined to be the angle between the
pion track and the spectator kaon track in the rest
frame of the \KstarzbEightNineTwo.
This angle depends on the spin of the intermediate resonance:
for \jBtoKKpi\ via the spin-1 \KstarzbEightNineTwo\ resonance, the
distribution of its cosine is quadratic.
For continuum events the distribution is approximately uniform.
At high \jH\ the final state pion has low momentum and
is difficult to reconstruct.
This causes a sharp drop-off in the efficiency
between $0.9$ and $1.0$.
The selection criterion $\jH < 0.9$ makes an unbinned
fit to the variable possible at a cost of losing $5 \%$
of signal events.

After the selection described above, the \Btokstarktokpi\ selection
efficiency is \sigEff.
In signal MC studies, the signal candidate is correctly reconstructed
$94 \%$ of the time.
The remaining candidates come from self-cross-feed (SCF) events
that stem from swapping one or more tracks from the true \B\ meson
decay with tracks from the rest of the event.

To identify backgrounds from \B\ meson decays the
selection criteria described above are applied to
the MC samples.
Using the efficiencies of the selection criteria
and world average branching fractions we find
that the largest expected contributions derive from
\btoc\ transitions and from charmless
3-body decays in which a kaon is misidentified as a
pion or vice versa.
The \btoc\ events, combinatoric in nature, are continuum-like
in most of the fit variables.
As such, the
contribution to
the fitted signal is small:
$\sim 1 \%$ of the number of \btoc\ events
expected to be present.
For charmless 3-body sources, however, the
contribution to
the signal yield---as a proportion of the number
of events present---is considerably larger.
This is particularly true of \BtoKKK\ and \BtoRhoK.
The decay \BtoKKK\ includes non-resonant and several intermediate
resonance states, including the narrow $\phi$ state.
In order to increase statistical precision, $\phi \Kp$
is considered separately from the rest of the $\Kp \Km \Kp$ final
state, which is modeled using the results of the Dalitz plot analysis
in~\cite{Garmash:2004wa}.
Since the $\phi$ is very narrow, its interference with the other
$\Kp \Km \Kp$ states can be neglected in this context.
The modes \BtoPhiK, \BtoKKK, and \BtoRhoK, are therefore included
as components of the fit.
This eliminates biases on the fitted signal yield
due to these channels.

The contribution to the signal yield due to all other
sources of \B\ meson background (including the \btoc\ modes
discussed above and numerous charmless modes) is estimated
from simulation at \nonKKpBBsub\ events---to which a
conservative $\pm100 \%$ uncertainty is assigned.
The uncertainty accounts for poorly known branching fractions and
simulation limitations.
Individually these sources contribute at a low level.
As such, rather than including components for each of them in
the fit, a correction to the fitted signal yield is made.

It is also necessary to consider backgrounds from
\B\ meson decays that have the same final state as
the signal mode.
MC studies of the \jBtoKKpi\ Dalitz plot show that the
only contribution that needs to be accounted for is
\BtokstarIIktokpi.
We use the LASS parameterization for the \KstarII\ lineshape,
which consists of the \KstarII\ resonance together with an
effective range non-resonant component~\cite{Aston:1987ir}.
We take $\BR(\KstarIItokpi)$ to be equal to
$\frac{2}{3} \times (93\pm10) \%$~\cite{Eidelman:2004wy}.
A maximum likelihood fit to three variables---\jM, \jD, and
\jF ---is performed in a region
of the \jKpi\ invariant mass spectrum between $1.048$
and $1.800 \gevcc$ in an analogous way to how we fit
the main signal (see below).
A $90 \%$ CL upper limit of $\BRulII$ is
placed on $\BR(\BtokstarIIk)$.
From simulation and the branching fraction $\BR(\BtokstarIIk)$ as obtained
above, we estimate---assuming zero interference---that
\BtokstarIIkPresent\ \BtokstarIIktokpi\ events will be present in the
region $0.744<\jI<1.048 \gevcc$, contributing \BtokstarIIkBias\ events
to the fitted \Btokstarktokpi\ signal yield.
The central value of this estimated contribution is calculated
from the central value of $\BR(\BtokstarIIk)$
while the uncertainty on the contribution covers
the contribution obtained from the upper limit on $\BR(\BtokstarIIk)$
and uncertainties in the parameterization
of the \KstarII\ lineshape.

A correction is applied to the fitted \Btokstarktokpi\ yield to account for
\BtokstarIIktokpi;
including a component in the fit is ineffectual since \BtokstarIIktokpi\ is
signal-like in the majority of the fit variables.

An unbinned extended maximum likelihood fit to the five
variables \jM, \jD, \jF, \jI, and \jH, is used to extract
the total number of
\Btokstarktokpi\ and continuum
background events.
The likelihood for the selected sample is given by the
product of the probability density functions (PDFs) for
each individual candidate, multiplied by the Poisson factor:
\begin{equation}
\label{likeEq}
\Like = \frac{1}{N!}\,e^{-N^\prime}\,(N^\prime)^N\,\prod_{i=1}^N {\cal P}_{i},
\end{equation}
where $N$ and $N^\prime$ are the number of observed and
expected events, respectively.
The PDF ${\cal P}_i$ for a given event $i$ is the sum of
the signal (S) and background (B) terms:
\begin{equation}
{\cal P}_{i} = N^{S} \left( (1 - f) \left( {\cal P}^{S} \right)_{i} + f \left( {\cal P}^{S}_{SCF} \right)_{i} \right)
+ \sum_{j=1}^{4} N^{B}_{j} \left( {\cal  P}^{B}_{j} \right)_{i},
\end{equation}
where $N^{S}$ and $N^{B}_{j}$ are the yields for
the signal component and the background components $j$,
and $f = 0.06$ is the fraction of SCF signal events
(treating true signal and SCF separately reduces
the correlation between \jM\ and \jH\ for signal from
$14 \%$ to $1 \%$).
The four background terms comprise the continuum
distribution and the three \B\ meson background modes
described above.
The PDF for each component is the product of the PDFs for
each of the fit input variables: ${\cal P} =
{\cal P}_{\jM}\,{\cal P}_{\jD}\,{\cal P}_{\jF}\,{\cal P}_{\jI}\,{\cal P}_{\jH}$.
Any correlations between the variables are such that
biases brought about in the fit are negligible and we
treat each PDF as independent and uncorrelated.

The PDF forms are presented in Table \ref{tab:PDFForms}.
The parameters of the signal and \B\ meson background PDFs are held
fixed to the MC values.
The parameters of the continuum PDFs are allowed to float except for
the endpoint of the \ARGUS\ function.
The signal and continuum yields are floated in the fit while the three
\B\ meson background yields are fixed to their MC expectations.

\begin{table}[bt]
\caption[PDF forms.]{PDFs are described by one or more of the following functions: \ARGUS~\cite{Albrecht:1990am} (A),
Breit-Wigner (BW), Crystal Ball~\cite{CBRef} (CB), double Gaussian (DG), \expFunc---where $P$ is a polynomial in the fit
variable and $C$ is a real scalar (E), Gaussian (G), linear (L), one-dimensional non-parametric~\cite{Cranmer:2000du} (N), quadratic
(Q), Voigtian---a Gaussian convolved with a Breit-Wigner (V).}
\label{tab:PDFForms}
\begin{center}
\resizebox{\columnwidth}{!}{
\begin{tabular}{lcccccc}
\hline
\multirow{2}{*}{{\bf Component}}   &  \hspace{2ex}  &  \multicolumn{5}{c}{{\bf PDF variable}}                                                     \\
                                   &  \hspace{2ex}  &  \jMb             &  \jDb             &  \jFb          &  \jIb              &  \jHb         \\
\hline
\multicolumn{7}{l}{Signal}                                                                                                                        \\
\hspace{1em}Truth-matched          &  \hspace{2ex}  &  CB               &  DG               &  DG            &  BW                &  E            \\
\hspace{1em}SCF                    &  \hspace{2ex}  &  CB               &  L                &  DG            &  G$+$L             &  \efc E \efc  \\
\hline
\multicolumn{7}{l}{Background}                                                                                                                    \\
\hspace{1em}Continuum              &  \hspace{2ex}  &  A                &  L                &  DG            &  \efc BW$+$L \efc  &  Q            \\
\hspace{1em}\BtoPhiK               &  \hspace{2ex}  &  CB               &  G$+$L            &  \efc DG \efc  &  V$+$E             &  E            \\
\hspace{1em}\BtoKKK                &  \hspace{2ex}  &  \efc A$+$G \efc  &  \efc G$+$L \efc  &  DG            &  Q                 &  N            \\
\hspace{1em}\BtoRhoK               &  \hspace{2ex}  &  CB               &  G$+$L            &  G             &  G$+$L             &  E            \\
\hline
\end{tabular}
}
\end{center}
\end{table}

Individual contributions to the systematic uncertainty are summarized
in Table \ref{tab:systematics}.
The systematic uncertainties that arise from fixing PDF parameters
for the signal and \B\ meson background components
are estimated by varying these parameters, one at a time.
Correlated parameters in the relevant PDF are adjusted accordingly
and the maximum likelihood fit is repeated with the shift in
the signal yield taken to be the systematic uncertainty.
The parameters are varied either by the $1\sigma$ uncertainties obtained
when evaluating them from MC or such that we account for any discrepancy
observed between data and MC (whichever is larger).
Such discrepancies are identified with the calibration channel
\jimBtoDpiCalib, which has a topology similar to the signal and
a much higher branching fraction.
For ${\cal P}^{S}_{\jH}$, rather than varying the PDF parameters,
a second-order polynomial is used.
This is the expected shape when neglecting detector and
reconstruction effects.
For the non-parametric PDF, its smoothness is varied.
The positive and negative shifts for each varied PDF parameter/shape
are added separately in quadrature.
The same procedure is used for the fixed yields of the \B\ meson
background modes and for the SCF fraction, $f$, which is varied
by $\pm20 \%$ (relative).
The uncertainty due to fixing the \ARGUS\ endpoint for the
continuum background \jM\ PDF is found by floating this parameter and
observing the shift in the signal yield; this shift is found to be less than
a hundredth of an event.
The systematic uncertainties associated with the subtraction
of events from the fitted signal yield due to
\BtokstarIIktokpi\ and other non-\jKKpi-final-state
\B\ meson decays have been discussed above.

The remaining systematic uncertainty on the signal yield
is due to possible interference between final states.
Several thousand MC datasets are produced each containing
\Btokstarktokpi\ and \BtokstarIIktokpi\ events that are
generated according to their Breit-Wigner and LASS lineshapes,
respectively.
Interference between the two modes at the amplitude level is
modeled.
The relative magnitudes and phases of the two contributing
amplitudes are varied randomly between datasets, but the numbers
of events present in the regions $0.744<\jI<1.048 \gevcc$ and
$1.048<\jI<1.800 \gevcc$ are the same for each dataset and are
equal to the numbers we observe in data.
The fractional systematic uncertainty is taken to be twice
the standard deviation of the distribution of the fraction
$f_{892}$ divided by its average and is found to be $8 \%$.
For a generated dataset, $f_{892}$ is the modulus squared
of the amplitude of the \Btokstarktokpi\ mode integrated over the full
Dalitz plot, divided by the modulus squared of the sum of
the amplitudes of the \Btokstarktokpi\ and \BtokstarIIktokpi\ modes integrated over the full Dalitz plot.

The uncertainties on reconstruction and selection criteria
efficiencies are evaluated by comparing efficiencies for MC
and data control samples.
A systematic uncertainty of $\pm1.4 \%$ per track added linearly
is taken for the particle identification efficiency.
A systematic uncertainty of $\pm0.8 \%$ on the tracking efficiency
is applied for each charged track, added linearly.
The systematic uncertainty on the efficiency of the selection
criteria is found to be \effSystErr.
The systematic uncertainty on the total number of \B\ events is $\pm 1.1 \%$.

\renewcommand\baselinestretch{1.3}
\begin{table}[bt]
\caption[Breakdown of systematic uncertainties.]{Breakdown of systematic uncertainties.}
\label{tab:systematics}
\begin{center}
\begin{tabular}{lc}
\hline
{\bf Systematic effect}                                                &  {\bf Uncertainty}                       \\
\hline
\multicolumn{2}{l}{Yield}                                                                                         \\
\hspace{2em}Fixed PDF parameters                                       &  {\normalsize{$^{+1.8}_{-1.1}$}} events  \\
\hspace{2em}Fixed SCF fraction                                         &  $\pm 0.7$ events                        \\
\hspace{2em}Fixed \B\ meson background yields in fit                   &  {\normalsize{$^{+0.4}_{-0.6}$}} events  \\
\hspace{2em}\B\ meson background contribution                          &  ~                                       \\
\hspace{4em}Non-\jKKpi-final-state                                     &  $\pm 3.2$ events                        \\
\hspace{4em}\jKKpi-final-state                                         &  {\normalsize{$^{+2.0}_{-5.8}$}} events  \\
\hspace{2em}Final state interference                                   &  $\pm 2.0$ events                        \\
\hline
\multicolumn{2}{l}{Reconstruction and selection criteria efficiency (RSC)}                                        \\
\hspace{2em}Tracking                                                   &  $\pm 2.4 \%$                            \\
\hspace{2em}Particle identification                                    &  $\pm 4.2 \%$                            \\
\hspace{2em}Selection criteria                                         &  $\pm 13.0 \%$                           \\
\hline
Yield total                                                            &  {\normalsize{$^{+4.7}_{-7.1}$}} events  \\
RSC total                                                              &  $\pm 13.8 \%$                           \\
Total number of \B\ events                                             &  $\pm 1.1 \%$                            \\
\hline
Total systematic uncertainty ($\times 10^{-6}$) on                     &  \jmrTwo{$\pm 0.2$}                      \\
$\BR(\Btokstark)$                                                      &                                          \\
\hline
\end{tabular}
\end{center}
\end{table}
\renewcommand\baselinestretch{1.0}

\begin{figure}[tb]
\begin{tabular}{@{}c@{}@{}c@{}}
\epsfig{file=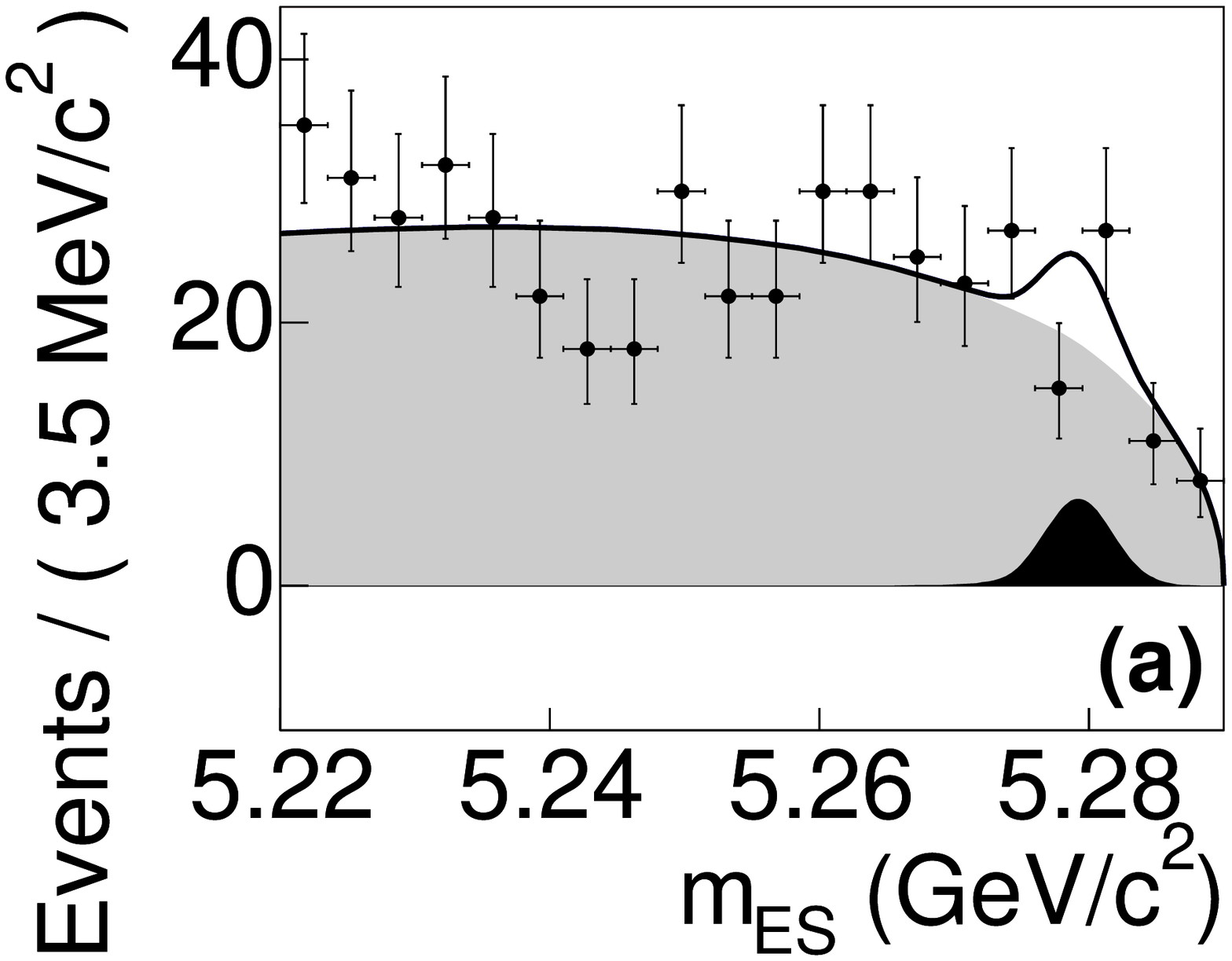,width=0.5\columnwidth}&
\epsfig{file=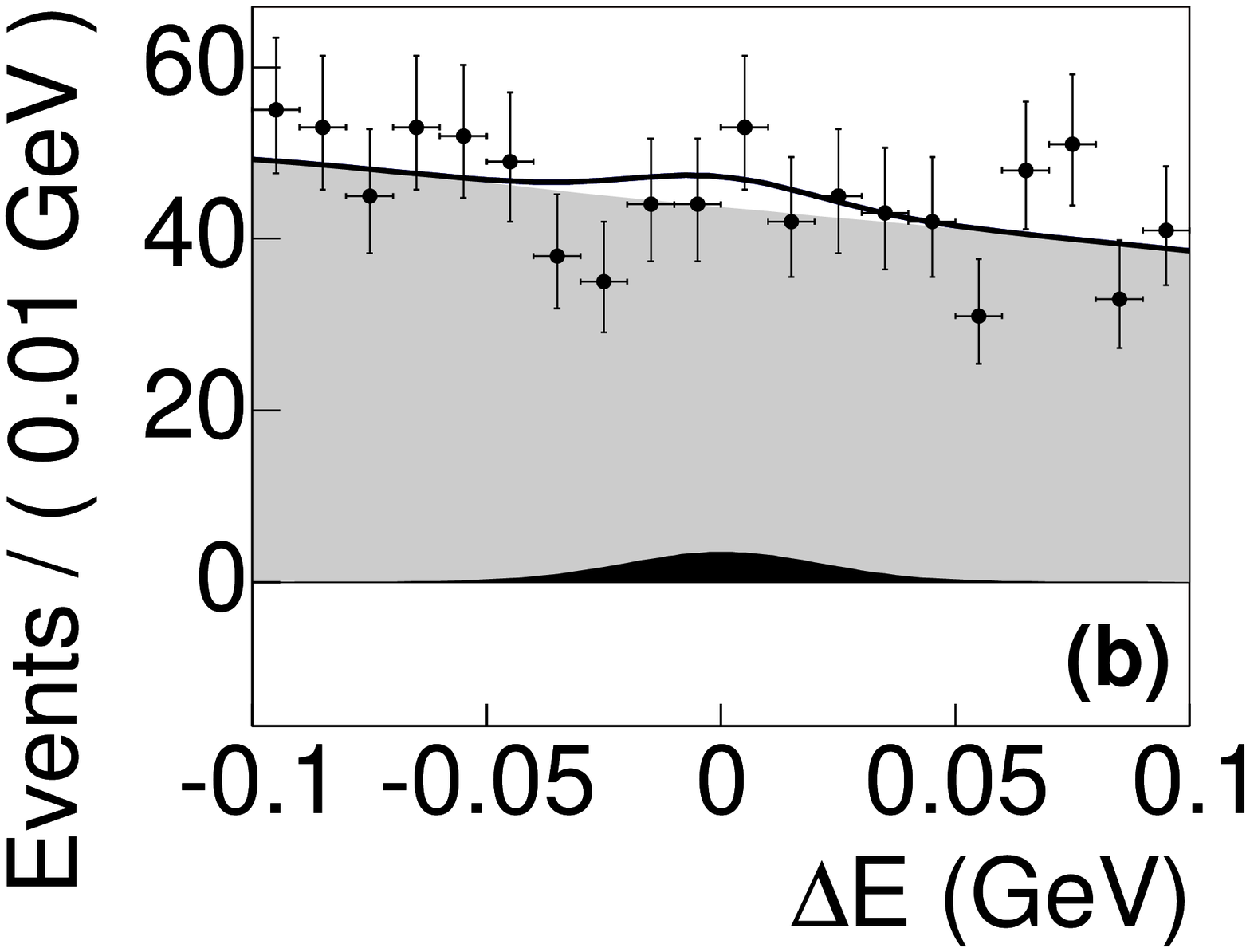,width=0.5\columnwidth}\\
\epsfig{file=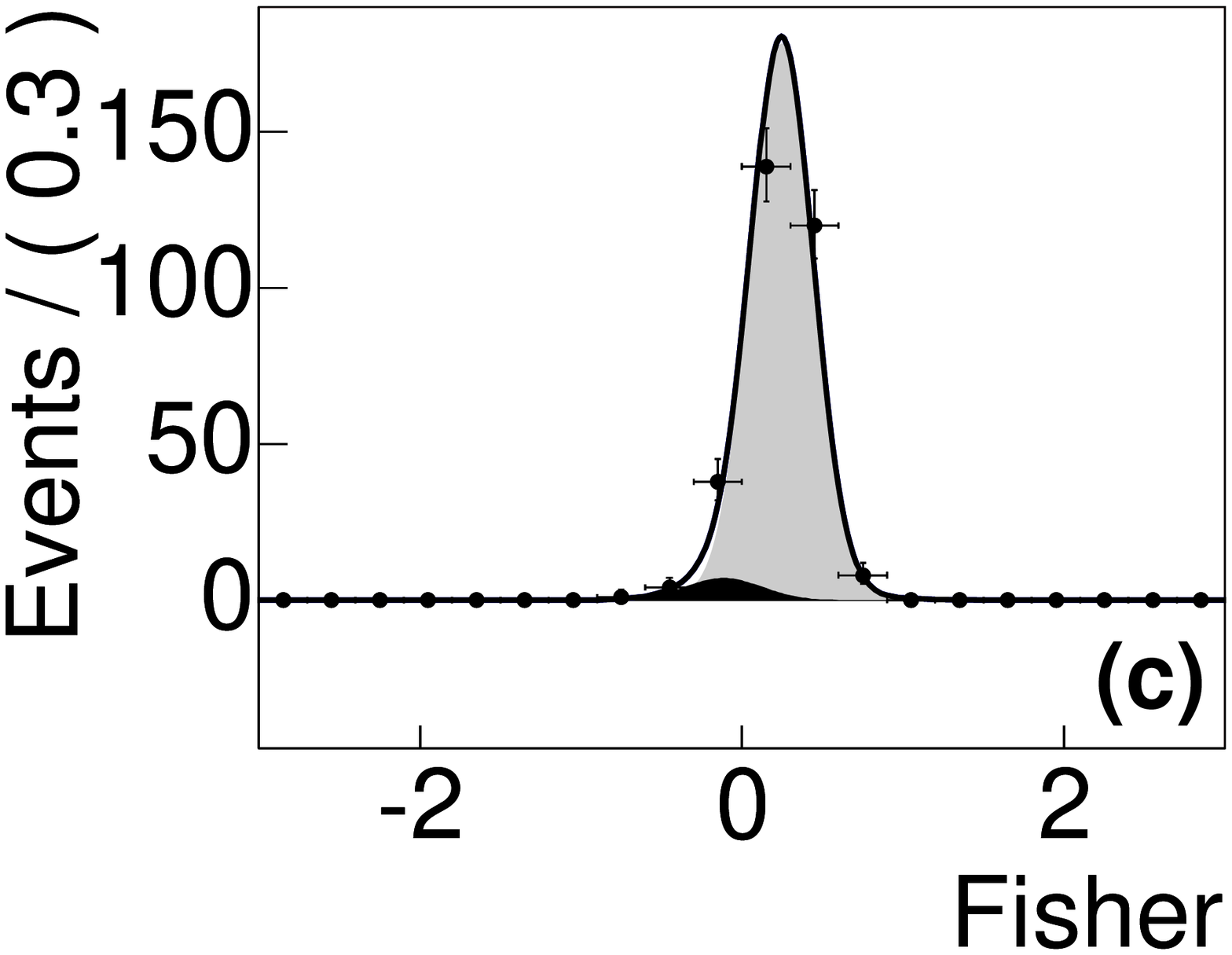,width=0.5\columnwidth}&
\epsfig{file=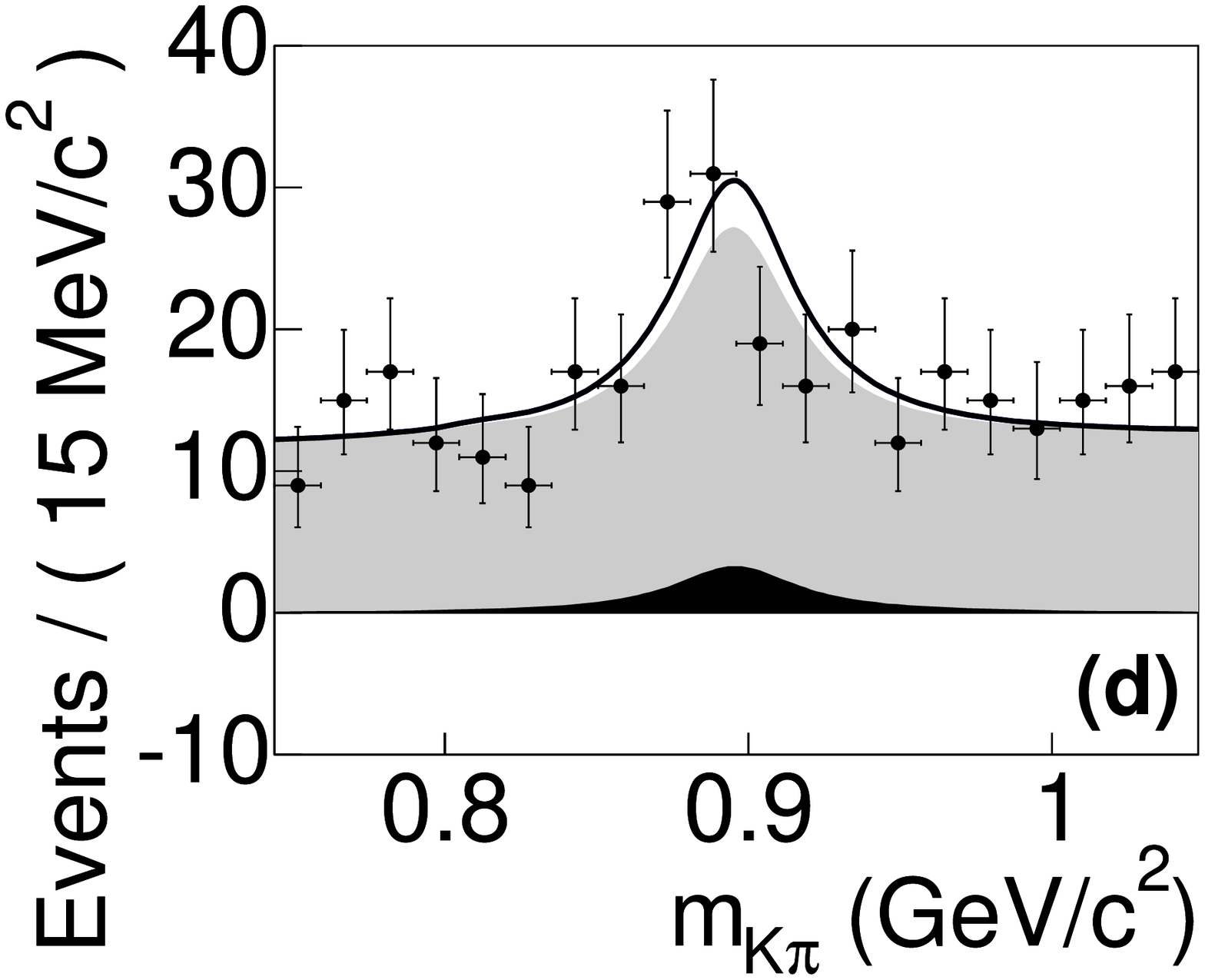,width=0.5\columnwidth}\\
\epsfig{file=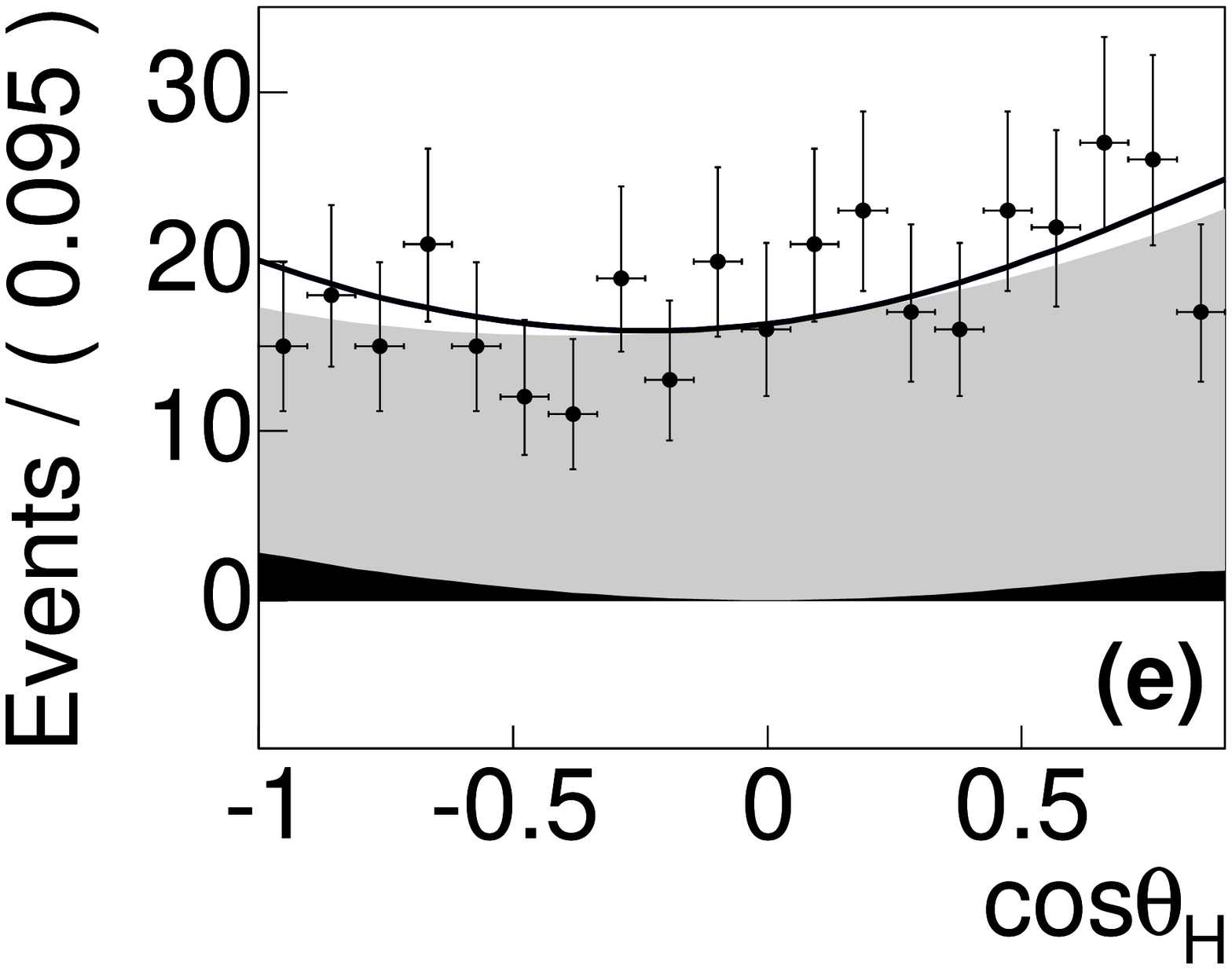,width=0.5\columnwidth}&
\epsfig{file=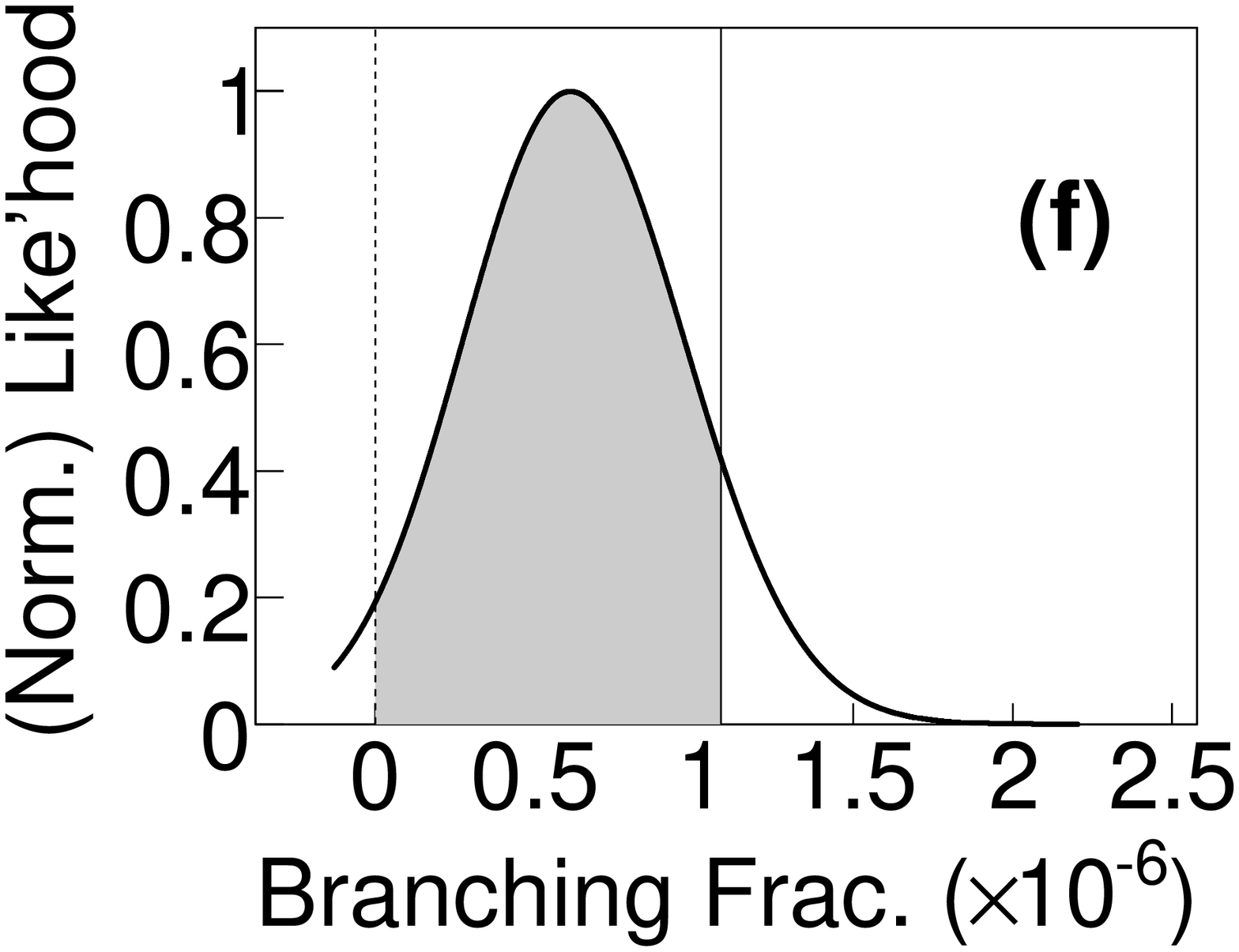,width=0.5\columnwidth}\\
\end{tabular}
\caption[Fit variable distributions and likelihood curve.]
{
(a)-(e):
distributions of \jM, \jD, \jF, \jI, and \jH.
The points with uncertainties show the data.
The curves show projections of the maximum likelihood fit.
A selection requirement on the likelihood ratio has been applied
as described in the text.
The black, solid curve (no filling) shows the sum of all fitted components.
The curve with gray filling shows the sum of all background components.
The curve with black filling shows the signal component;
(f):
likelihood (modified to account for systematic uncertainties, and normalized)
as a function of $\BR(\Btokstark)$.
The shaded area represents $90 \%$ of the total area under
the curve in the positive branching fraction region.}
\label{fig:results}
\end{figure}

A total of \numEvts\ events are fitted.
The numbers of \BtoPhiK, \BtoKKK, and \BtoRhoK\ events
expected in this sample, estimated from simulation, are
\BtoPhiKYld, \BtoKKKYld, and \BtoRhoYld, respectively.
The yields for these \B\ meson background components
are fixed at the central values.
The raw signal yield extracted from the fit is \rawYield\ events,
of which \BBbkgdSub\ are estimated to be \B\ meson
background events.
The
number of true signal events present
in the on-resonance data sample is therefore \sigYield.
The statistical significance of the result in the absence
of systematic uncertainties, defined as the square root of
the difference between the value of $- 2 \ln \Like$ for zero
signal events and at its minimum, is \signif.
Accounting for systematic uncertainties, this significance
is reduced to \signifWithSyst.
The number of signal events is divided by the product of the signal
efficiency and the total number of \B\ events
to give the branching fraction
$\BR(\Btokstark) = \BRmeasMoreDet$.
Since the signal is not significant, we
place an upper limit on this measurement at the $90 \%$
CL: $\BR(\Btokstark) < \BRul$.
The likelihood function defined in Eq.~(\ref{likeEq})
is modified to incorporate systematic uncertainties
through convolution with a bifurcated Gaussian whose
standard deviations are set to the (asymmetric) total
systematic uncertainties described above.
The $90 \%$ CL upper limit is then defined to be the
value of the branching fraction $\BR(\Btokstark)$
(which corresponds to a particular value of $N^{S}$)
below which lies $90 \%$ of the total integral of the
modified likelihood function in the positive branching
fraction region.
This is illustrated in \figureref{results} (bottom row, right).

The results of the fit to \Btokstarktokpi\ are illustrated in \figureref{results}.
The plots are enhanced in signal by selecting only those events that
exceed an optimized threshold for the likelihood ratio
$R = N^{S}\,{\cal P}^{S}/(N^{S}\,{\cal P}^{S}+\sum_{j=1}^{4}N^{B}_{j}\,{\cal  P}^{B}_{j})$
where $N$ are the central values of the yields and ${\cal P}$ are the PDFs
with the projected variable integrated out.

In conclusion, we have reduced the $90 \%$ CL upper
limit on the branching fraction for the decay \Btokstark\ from
\cleoRes~\cite{Jessop:2000bv} to \BRul.
The central value has been measured to be \BRmeas\ with a
significance of \signifWithSyst, consistent with the
predictions of~\cite{Fleischer:2004rnAndFleischer:2005gj},~\cite{Chiang:2003pm},
and~\cite{Guo:2006uq}.
This measurement can be used to determine an upper bound on
\deltaS.
The technique described in Sec.~VI of~\cite{Aubert:2006wu}
is used, together with Eq.~(29) of~\cite{Grossman:2003qp} and
the findings of the \BtoPhipi\ and \BtoPhiK\ analyses described
in~\cite{Aubert:2006nn}
and~\cite{Aubert:2006nu,Garmash:2004wa,Briere:2001ue,Acosta:2005eu},
respectively, to place a $90 \%$ CL upper bound of \deltaSBound\ on
$|\deltaS|$.
Systematic uncertainties on the branching fractions used to
determine this bound are accounted for.
The bound presented here is significantly more restrictive than the bounds of
$\approx 0.4$ found in~\cite{Aubert:2006wu,Aubert:2006fy}.
We note that these latter bounds are based on fewer
theoretical assumptions (see~\cite{Grossman:2003qp}).
We have also placed an upper limit of \BRulII\ on the previously
unmeasured branching fraction of the decay \BtokstarIIk.

We are grateful for the excellent luminosity and machine conditions
provided by our \pep2\ colleagues, 
and for the substantial dedicated effort from
the computing organizations that support \babar.
The collaborating institutions wish to thank 
SLAC for its support and kind hospitality. 
This work is supported by
DOE
and NSF (USA),
NSERC (Canada),
CEA and
CNRS-IN2P3
(France),
BMBF and DFG
(Germany),
INFN (Italy),
FOM (The Netherlands),
NFR (Norway),
MIST (Russia),
MEC (Spain), and
STFC (United Kingdom). 
Individuals have received support from the
Marie Curie EIF (European Union) and
the A.~P.~Sloan Foundation.

\bibliography{KstK_References}
\addcontentsline{toc}{section}{\numberline{}References}
\bibliographystyle{apsrev}

\end{document}